\documentclass{gGAF2e}
\usepackage[dvipsnames]{xcolor}

\def\der#1#2{{\partial #1\over \partial #2}}

\newcommand{\be}{\begin{equation}}
\newcommand{\ee}{\end{equation}}

\def\bea{\begin{eqnarray}}

\def\eea{\end{eqnarray}}

\def\bse{\begin{subequations}}

\def\ese{\end{subequations}}

\def\bsea{\begin{subeqnarray}}

\def\esea{\end{subeqnarray}}

\def\({\left (}

\def\){\right )}

\def\[{\left [}

\def\]{\right ]}

\def\<{\left <}

\def\>{\right >}

\begin{document}

\jvol{00} \jnum{00} \jyear{2012} 

\markboth{\rm R. YELLIN-BERGOVOY ET AL}{\rm GEOPHYSICAL \&  ASTROPHYSICAL FLUID DYNAMICS}

\title{A minimal model for vertical shear instability in protoplanetary accretion disks}

\author{Ron Yellin-Bergovoy${\dag}$$^{\ast}$\thanks{$^\ast$Ron Yellin-Bergovoy. Email: yellinr@tau.ac.il},  Orkan M. Umurhan${\ddag}$${\diamond}$, and Eyal Heifetz${\dag}$\\\vspace{6pt}  ${\dag}$The Porter School of Environmental and Earth Sciences, Tel-Aviv University, Tel-Aviv, Israel\\
${\ddag}$ SETI Institute, Mountain View, CA 94035 USA\\
${\diamond}$NASA Ames Research Center, Division of Space Sciences, Planetary Systems Branch, Moffett Field, CA, 94035 USA\\\vspace{6pt}\received{v4.4 released October 2012} }

\maketitle

\begin{abstract}

The Vertical Shear Instability is an axisymmetric effect suggested to drive turbulence in the magnetically inactive zones of protoplanetary accretion disks. Here we examine its physical mechanism in analytically tractable  ``minimal models" in three settings that include
a uniform density fluid, a stratified atmosphere, and a shearing-box section of a protoplanetary disk.
Each of these analyses show that the vertical shear instability's essence is similar to the slantwise convective symmetric instability in the mid-latitude Earth atmosphere, in the presence of vertical shear of the baroclinic jet stream, as well as mixing in the top layers of the Gulf Stream.
We show that in order to obtain instability the fluid parcels' slope should exceed the slope of the mean absolute momentum in the disk radial-vertical plane. We provide a detailed and mutually self-consistent physical explanation from three perspectives: in terms of angular momentum conservation, as a dynamical interplay between a fluid's radial and azimuthal vorticity components, and from an energy perspective involving a generalised Solberg-H{\o}iland Rayleigh condition. Furthermore, we explain why anelastic dynamics yields oscillatory unstable modes and isolate the oscillation mechanism from the instability one.


\end{abstract}
\section{Introduction}
Fifty years of spectroscopic observations of young stellar objects (YSO) reveal them to exhibit strong UV excess indicative of some type of anomalous torque inducing process driving transport of mass and angular momentum within the YSO's surrounding accretion disk.   There are a variety of processes that might give rise to this phenomenon, but highest on the list are winds and turbulence.  In the latter case, the source of turbulence depends upon the temperature and ionization state of the accretion disk gas.  Around relatively cold planet forming  ``protoplanetary" disks, where the degree of ionization is practically zero, identifying candidate turbulence generating mechanisms had remained stubbornly elusive until the last 5-10 years where three viable linear instability processes have been identified \citep[for detailed discussion of all mechanisms and their context in planet forming disks, see,][]{2019PASP..131g2001L}.  Of these, the vertical shear instability (``VSI" hereafter) \citep[]{1998MNRAS.294..399U, 2004A&A...426..755A, 2013MNRAS.435.2610N, 2014A&A...572A..77S} appears to be most relevant inside the 5-50 AU ``Ohmic Zone" (OZ hereafter)\footnote{i.e., Ohmic zones are regions in the disk where Ohmic resistivity is so large that MHD effects are effectively inoperative.} of a protoplanetary disk like our own solar nebula \citep{2017A&A...605A..30M} -- the zone where the cores of the gas and ice giants, as well as the Kuiper Belt objects, were likely assembled from 20-100 km sized planetesimals. Turbulence affects the
way planetesimals are manufactured \citep[e.g.][]{Umurhan_etal_2020,Hartlep_etal_2020,Chen_Lin_2020} and given that the VSI appears likely to be central in driving turbulence in the outer solar system, understanding how it operates from a mechanistic view is essential for understanding how it interacts with planetesimal formation scenarios. 
\par
The VSI is the zero Prandtl number, rapid thermal relaxation time limit of the Goldreich-Schubert-Fricke instability
\citep[][``GSF" hereafter]{Goldreich_Schubert_1967,Fricke_1968},
which has been examined as a possible mixing-agent in radiative zones of differentialy rotating stars. 
\citet{knobloch1982stability} show that
the velocity displacement of unstable GSF modes are inclined, lying between the rotation axis and surfaces of constant angular momentum \citep[see also][]{latter2018vortices}.  \citet{knobloch1982stability} further explain that while buoyancy acts to stabilize these modes, sufficiently small enough radial scales can become GSF unstable by buoyancy suppression brought about by increased thermal diffusivity \citep[see also][]{Lin_Youdin_2015}.
VSI is essentially an instability of inertial modes that arises in a strongly rotating disk flow supporting a radial temperature gradient. This temperature profile induces a vertical shear in the azimuthal (Keplerian) flow of the gas around the central star.  This vertical shear arises in the same manner as does the thermal wind solution for the Earth.  Furthermore, the VSI appears to be analogous to the long-studied Symmetric Instability thought to be an important mixing-agent for mesoscale and sub-mesoscale dynamics in both the Earth's oceans (like the Gulf Stream) and atmosphere (mid-latitudes) \citep[e.g.,][and other studies cited within]{Thomas_etal_2013,Stamper_Taylor_2017}.
An important ingredient for the VSI's operation is that the thermal relaxation times are very short compared to the local orbital times \citep[nominally quantifying the frequency timescales of the unstable carrier inertial modes, see also,][]{2013MNRAS.435.2610N}.  In this sense, the instability is thermally driven and its energy is ultimately derived by the star's light and is stabilized with increased vertical stable stratification, \citet{Lin_Youdin_2015}. The exact location in a disk with conditions favorable for the VSI to dominate over other magnetic and/or purely hydrodynamic instabilities is still under debate, as these criteria depend open both the disk model and the thermodynamic response of the disk, both of which are still not fully understood \citep{ 2017A&A...605A..30M,pfeil2019mapping, 2019PASP..131g2001L}.
While VSI surface modes exhibit approximately similar -- if not slightly weaker -- growth rates compared to the body modes \citep{barker2015vertical}, it was argued in \citet{Umurhan_etal_2016} that if one were to project the energy of a nonlinearly developed state of the VSI onto the set of eigenmodes of the system (as, for example, observed in simulations), one would find that majority of that energy would project onto the lowest order body modes. In this work we therefore relate mostly to the body modes which are semi-global in the sense that they span the disk's full vertical extent, i.e., several pressure scale heights ($H$), while its radial scales are fractions of a scale height ($\sim \epsilon H$, where $\epsilon$ is the disk opening angle).

\par
Despite the VSI appearing to be central for OZ dynamics across large swaths of the solar nebula, its physical mechanism is not fully understood.  Identifying how the VSI works and understanding it in relation to other similar
processes in planetary atmospheres -- and perhaps establishing that it may be a member of a super-class of processes -- is in its own right a desirable physics-oriented intellectual end.
The instability is often explained in terms of the semi-quantitative Solberg-H{\o}iland  generalised Rayleigh-like condition
\citep{Umurhan_etal_2013,2014A&A...572A..77S,barker2015vertical,manger2018vortex}, which examines the net energy obtained when pairs of fluid parcels exchange positions -- which might place the process into the same Taylor-Couette family of instabilities.
\par
Our aim here is to clarify the mechanistic understanding of VSI by analyzing an analytically tractable minimal model (section 2). Similar to the Symmetric Instability mechanism \citep[e.g.,][]{holton2012introduction} in the Earth mid-latitudinal atmosphere, the VSI can be viewed from angular momentum (section 3), vorticity (section 4) and energetic perspectives (section 5). We show that all perspectives boil down to the condition that the fluid parcel's slope trajectory must exceed the mean absolute momentum slope in the disk's radial-vertical plane. Only then the gain in the azimuthal velocity via vertical advection of the mean shear overcomes the loss via the Coriolis deceleration due to radial motion.

\section{Minimal model for VSI}

Our starting point is the shearing box linearised asymptotic reduced set of equations, introduced by \cite{2013MNRAS.435.2610N}. For the detailed careful steps leading to the reduced model, the reader is kindly referred to the Appendix of that paper. The equation set is derived  by assuming that the spatial and temporal scales of motion are related to one another according to the following: temporal dynamics are given by ${\mathrm O}(1 / \epsilon \varOmega_{0}),$ radial dynamic scales $x$ are ${\mathrm O}(\epsilon^{2} R_{0}),$ and the vertical scales $z$ are on the scale height $H_{0}={\mathrm O}(\epsilon R_{0}),$ in which the small parameter $\epsilon \equiv H_{0} / R_{0}$ measures the disk opening angle.  Note specifically that in this reduced model the stability effect of vertical buoyancy is neglected.
Written in the slightly more general form the reduced model is governed by the following set of equations  
%
%
\bse
\begin{align}
0 =\,&\, C_x v - \der{\varPi }{x},\\
\dot{v} =\,&\, -C_y u + {\overline S}(z)w,\\
\dot{w} =\,&\, -\der{\varPi }{z},\\
\der{}{x}\[{\overline \rho}(z)u\] + \der{}{z}\[{\overline \rho}(z)w\] =\,&\, 0,
\end{align}
\ese
where dots over quantities denote $\upartial/\upartial t$.
Here we look at a box within the upper part of the disk where the Cartesian coordinates $(x,y,z)$ correspond approximately to the 
(radial, azimuthal, vertical) directions (figure  1(a)). $(u,v,w)$ are the corresponding perturbation velocity components and $\varPi $ is the  pressure perturbation, scaled by a reference density. $(C_x, C_y) = \varOmega_0(2,1/2)$ are the local $(x,y)$ components of the Coriolis parameter, taken to be constant within the box, where $\varOmega_0$ is the Keplerian rotation frequency at the box center so that $C_y$ is equal to the negative value of the radial Keplerian shear within the box. ${\overline S}(z) = -\upartial\overline V/ \upartial z > 0$ represents the negative vertical shear of the azimuthal mean flow (denoted by overbar) in the upper part of the disk (as the Keplerian flow decays away from the mid-plane) and ${\overline \rho}(z)$ is the mean density profile. We consider dynamics without variations in the azimuthal direction 
($\upartial{}/ \upartial y  \(\,\,\, \)= 0$).

 \begin{figure}[ht!]
 \begin{center}
\includegraphics[scale=0.83]{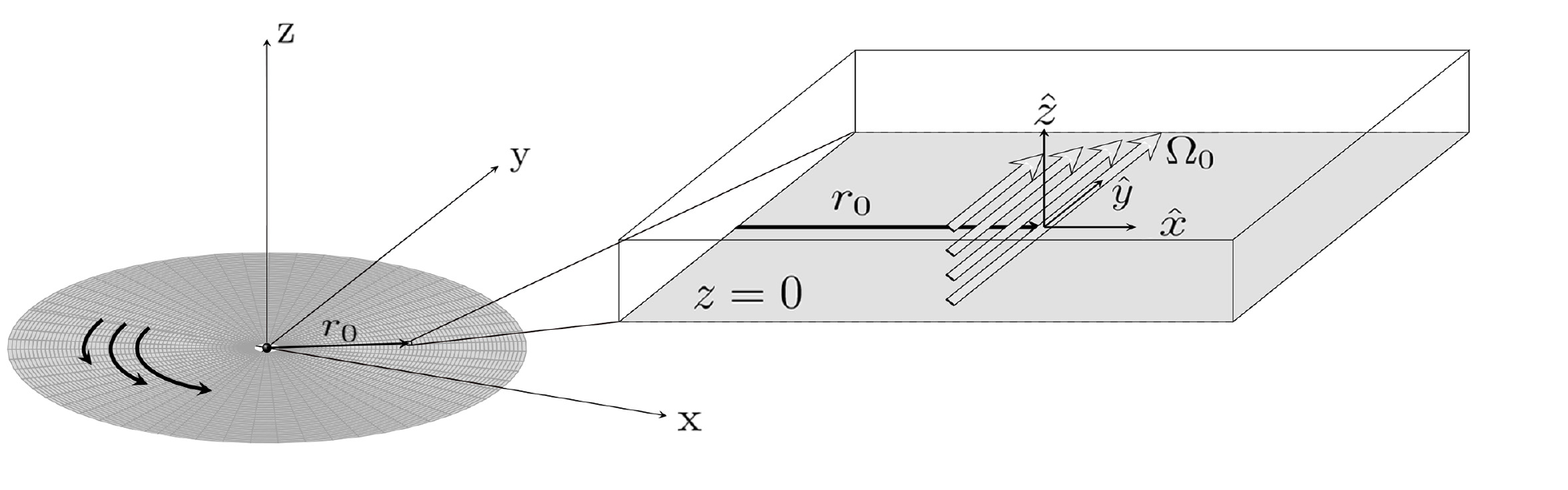}
\caption{Disk setup. Illustration of the shearing box approximation for Keplerian disks with diminishing azimuthal velocity from the mid-plane. The Cartesian coordinates $(x,y,z)$ correspond approximately to the (radial, azimuthal, vertical) directions.}
\end{center}
\end{figure}

 \begin{figure}[t!]
 \begin{center}
\includegraphics[scale=1.3]{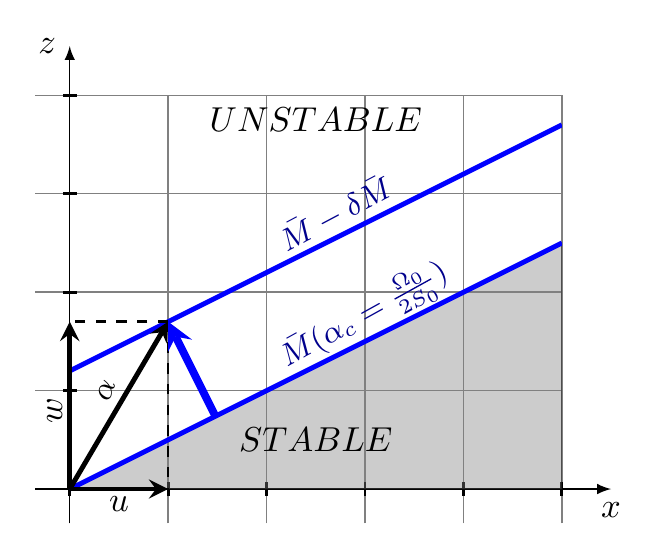}
\caption{Stability diagram on the $x-z$ (radial-vertical) plane. Instability is obtained when the mean particles trajectory slope, $\alpha = w/u$ (black line), exceeds the critical slope $\alpha_c =\varOmega_0/ (2S_0)$, of constant azimuthal absolute momentum (blue lines). The wedge of stability (between zero slope to the absolute momentum slope) is shaded in grey.   The blue arrow corresponds to the advection of the mean absolute momentum by the perturbation velocity. (Colour online)}
\end{center}
\end{figure}

The careful scale analysis of \cite{2013MNRAS.435.2610N} suggests the approximation in which the azimuthal perturbation velocity, $v$, is in geostrophic balance (1a). $v$ can change both due to radial and vertical motions (1b). Inner radial flow ($u<0$) generates azimuthal acceleration due to the Coriolis force and upward motion ($w>0$) advects large Keplerian azimuthal flow from the mid-plane. Equation (1c) states that vertical acceleration results from the vertical pressure gradient deviation from the mean hydrostatic balance. Finally, the continuity equation (1d) is assumed anelastic where only the vertical variations in the mean density are taking into account.

The simplest solution of (1),  admitting the simplest instance of the VSI, is obtained when both ${\overline \rho}$ and ${\overline S}$ are taken as constants 
$({\rho}_0 , {S}_0)$. Straightforward manipulations reduces the equation set (1) into a single constant coefficient partial differential equation for the pressure perturbation
\be
\der{^2 \ddot \varPi }{x^2} = -C_x\(S_0{\upartial {^2 \varPi }\over \upartial {x} \upartial {z}} + C_y \der{^2 \varPi }{z^2} \).
\ee
Substituting a plane-wave solution of the form of $\exp[{\mathrm i} (kx +mz -\omega t)]$ into the above yields the dispersion relation
\be
\omega^2 = {S_0 C_x \over \alpha ^2}\(  \alpha_c - \alpha \),
\ee
with $\alpha \equiv - k/m$. Equation (1d) indicates then that $\alpha= w/u$ is the fluid parcel slope in the $(x,z)$ plane. Hence, instability is obtained when:
\be
{w\over u} = \alpha >  \alpha_c = \dfrac{C_y}{S_0} > 0\, ,
\ee
i.e.,  when the parcel trajectory slopes are tilted upward and outward with a tilt that is larger than the critical slope of $C_y / S_0 = \varOmega_0/2 S_0$, which is the ratio between the absolute values of the azimuthal Keplerian shear and the vertical mean shear (figure 1(b))\footnote{From reflection symmetry, this condition holds as well for the lower part of the disk where $S_0$ is generally negative.}
This critical slope is the slope of the mean azimuthal absolute momentum surfaces, as explained in the following subsection.

\section{Absolute momentum perspective}

Denote the axi-symmetric dynamic total azimuthal velocity (deviating from the basic Keplerian disk flow) as  
$V(x,z)={\overline V}(z) +v(x,z)$, then the linearized momentum equation in the radial direction (1b) reads
\be
{\mathrm D \over \mathrm D t}V = -C_y u = -C_y {\mathrm D \over \mathrm D t}x\, ,
\ee
where ${\mathrm D}/ {\mathrm D} t= \upartial{}/{\upartial t} +{\bm u}\bm\cdot\bm\nabla$ is the material time derivative. Define $M \equiv V + C_y x$ as the absolute azimuthal linear  momentum (per unit mass),
equation (5) results then from the material conservation of $M$
\be
{\mathrm D \over \mathrm D t}M=0\, .
\ee

Define the mean absolute momentum as ${\overline M}(x,z) = {\overline V}(z) + C_y x$, and assess variations of its constituent terms along surfaces of constant ${\overline M}$. Thus
\be
\delta {\overline M} = 0 = \der{\overline M}{x} \delta x+  \der{\overline M}{z}\delta z = C_y \delta x - S_0 \delta z
\ee
yields the critical slope
\be
\({\delta z \over \delta x}\)_{\overline M} = \alpha_c\, .
\ee
Furthermore,  writing 
$M = {\overline M} +	{Green}{ \textsf{m}}$, then the perturbation azimuthal absolute momentum is simply $\textsf{m}=v$. Equation (1b) is then the linearized version of (6), namely
\be
\dot{v} = \dot{ \textsf{m}}=  -{\bm u}\bm\cdot\bm\nabla{\overline M} = -\(u\der{\overline M}{x} + w\der{\overline M}{z}\) = -C_y u + S_0 w =\(-{\alpha_c \over \alpha} +1 \)S_0 w,
\ee
where for the plane-wave solution (1c) becomes
\be
\dot{w} = {C_x \over \alpha}v.
\ee
Hence, mutual amplification between $v$ and $w$ is obtained when the gain in the azimuthal velocity $v$, via vertical advection of the mean shear (the term $S_0 w$ in (9)), overcomes its loss via the Coriolis deceleration due to radial motion (the term $-C_y u$). This is equivalent to negative advection of the mean absolute momentum by the perturbation (figure  2) as fluid parcels from an initial absolute momentum surface of ${\overline M}$, are displaced to the absolute momentum surface of ${\overline M}-\delta{\overline M}$, thus generating a positive anomaly $m$. We note as well that for instability the three components of the velocity should have the same sign (recall that $u= w/\alpha$ and $\alpha > 0$). 

Note that for an anomalous case where the vertical shear is reversed (${\overline S}<0$), the slope of ${\overline M}$ remains as in figure 2, however now $\upartial\overline M/ \upartial z > 0$. In this case  (9) and (10) indicate that instability may be achieved for negative parcel's slope ($\alpha <0$), under the condition $|S| > C_y/|\alpha|$.

\section{Vorticity perspective}

A different perspective on the amplification mechanism may be obtained when considering the interplay between the perturbation vorticity components in the $x$ direction, 
$\omega_x = -\upartial{v}/\upartial{z}$ (recall that $\upartial{}/\upartial{y}=0$) and in the $y$ direction, $\omega_y = \upartial{u}/\upartial{z} - \upartial{w}/\upartial{x}$. We first note that equations (1a,c)
yield 
\be
\der{\dot w}{x} = C_x \omega_x .
\ee   
In figure 3(a), it is shown how the geostrophic balance in (1a) leaves the positive (negative) pressure anomaly to the right (left) of $v$. Thus, negative vertical shear of the geostrophic velocity (positive $\omega_x$) imposes vertical pressure gradient structure which is pointing upward (downward) to the right (left) of the shear and therefore accelerates the vertical velocity in the opposite directions of the vertical pressure gradients.  Since $u$ is tied to $w$ by continuity,  together with the assumption of plane-wave solutions, equation (11) may be expressed in terms of the interplay of component vorticities
\be
{\dot \omega}_y = -C_x \(1+ \alpha^{-2}\)\omega_x.
\ee
On the other hand, we have 
\be
{\dot \omega}_x = C_y \der{u}{z} - S_0 \der{w}{z}.
\ee
The Coriolis acceleration to the right explains how the shear ($\upartial{u}/\upartial{z}$) generates $\omega_x$ (figure 3(b)), whereas the vertical advection of the mean shear by the vertical perturbation velocity explains how vertical convergence ($-\upartial{w}/\upartial{z}$) generates as well $\omega_x$ (figure 3(c)). Using continuity again we may write (13) for the plane-wave solution as
\be
{\dot \omega}_x =  - S_0 \({ \alpha -  \alpha_c \over 1+ \alpha^{2}}\)\omega_y .
\ee

\begin{figure}[ht!]
\begin{center}
\subfigure[]{
\resizebox*{8cm}{!}%
{\includegraphics{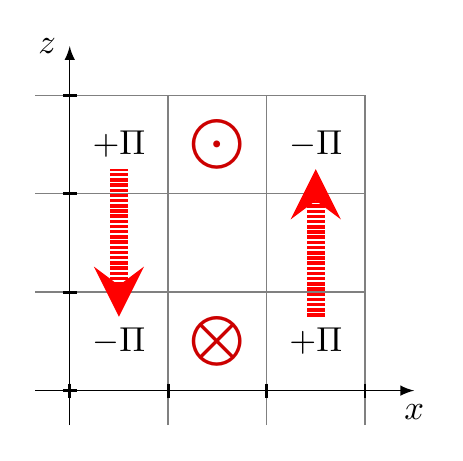}}}%
\subfigure[]{
\resizebox*{8cm}{!}%
{\includegraphics{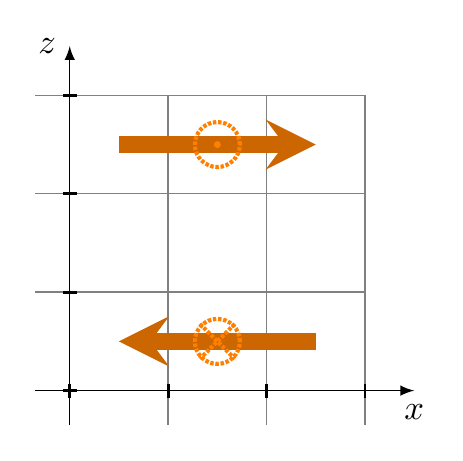}}}%
\vspace{0.5cm}
\subfigure[]{
\resizebox*{8cm}{!}%
{\includegraphics{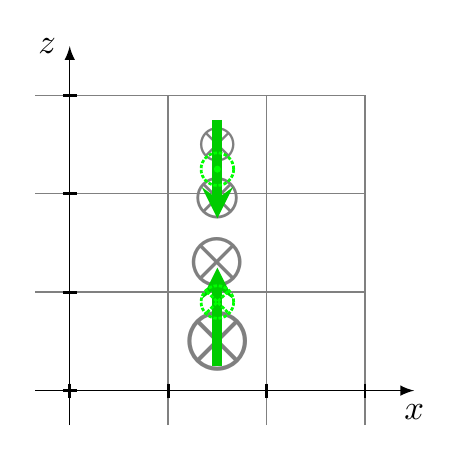}}}%
\caption{Different mechanisms for the generation of the different vorticity terms (see details in the main text). (a). Negative azimuthal vertical shear $-\upartial{v}/\upartial{z}$ generates positive vertical radial shear $ \upartial{w}/\upartial{x}$.  (b) Radial vertical shear $\upartial{u}/\upartial{z}$ generates negative azimuthal vertical shear $-\upartial{v}/\upartial{z}$. (c). Vertical convergence $-\upartial{w}/\upartial{z}$ generates as well negative azimuthal vertical shear $-\upartial{v}/\upartial{z}$.
In all sub-figures solid line arrows represent the generating mechanisms and dashed or dot arrows represent the responding dynamics. (Colour online)}
\end{center}
\end{figure}

It is clear from (12) and (14) that mutual amplification between the two vorticity components can be obtained only if $\alpha >  \alpha_c$ and when $\omega_x$ and $\omega_y$ are of opposite sign. Such scenario is demonstrated in figure 4(a) for the case where  $\omega_x < 0$ and $\omega_y >0$ . The pressure structure, resulting from the geostrophic balance associated with $-\omega_x$, amplifies $\omega_y$ according to figure 3(a) and (12). In turn, however, $\omega_y$ acts to diminish $\omega_x$ via the term $C_y (\upartial{u}/\upartial{z})$ (figure 4(b)), as described in figure 3(b). Thus, in order to maintain mutual amplification a strong enough vertical divergence is required (figure 4(c)) so that $|S_0 (\upartial{w}/\upartial{z})| >  |C_y (\upartial{u}/\upartial{z})|$, and for plane-waves this condition is indeed satisfied for $\alpha >  \alpha_c$. As a result, unstable eddies in the $(x,z)$ plane are elongated ellipses with a tilt that is steeper than the absolute momentum isolines slope (figure 4(d)). 
Since $\omega_z = \alpha \omega_x$, similar reasoning leads to unstable interplay between $\omega_y$ and $\omega_z$.

\begin{figure}[ht!]
\begin{center}
\subfigure[]{
\resizebox*{8cm}{!}%
{\includegraphics{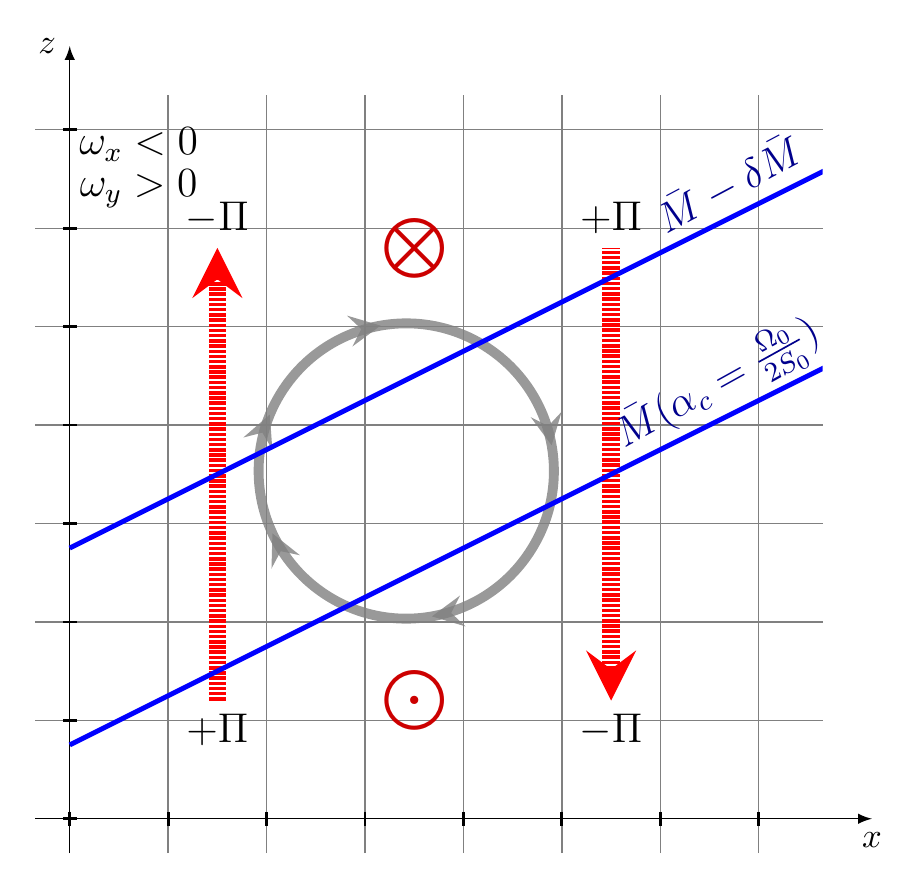}}}%
\subfigure[]{
\resizebox*{8cm}{!}%
{\includegraphics{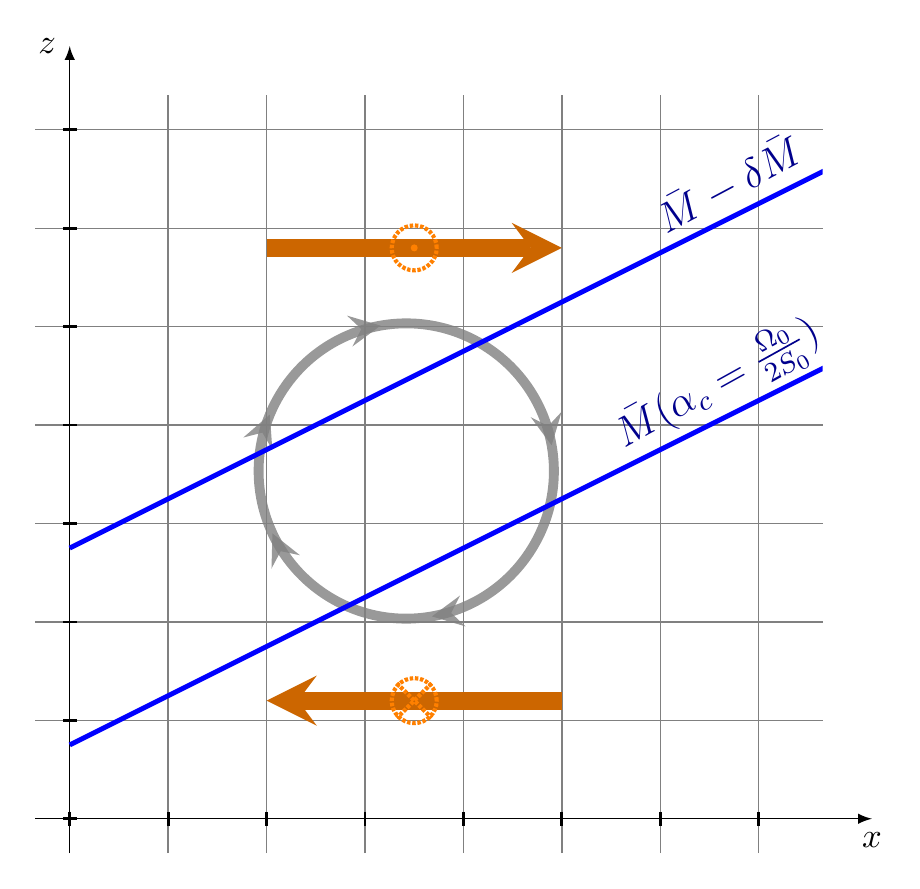}}}%
\vspace{0.5cm}
\subfigure[]{
\resizebox*{8cm}{!}%
{\includegraphics{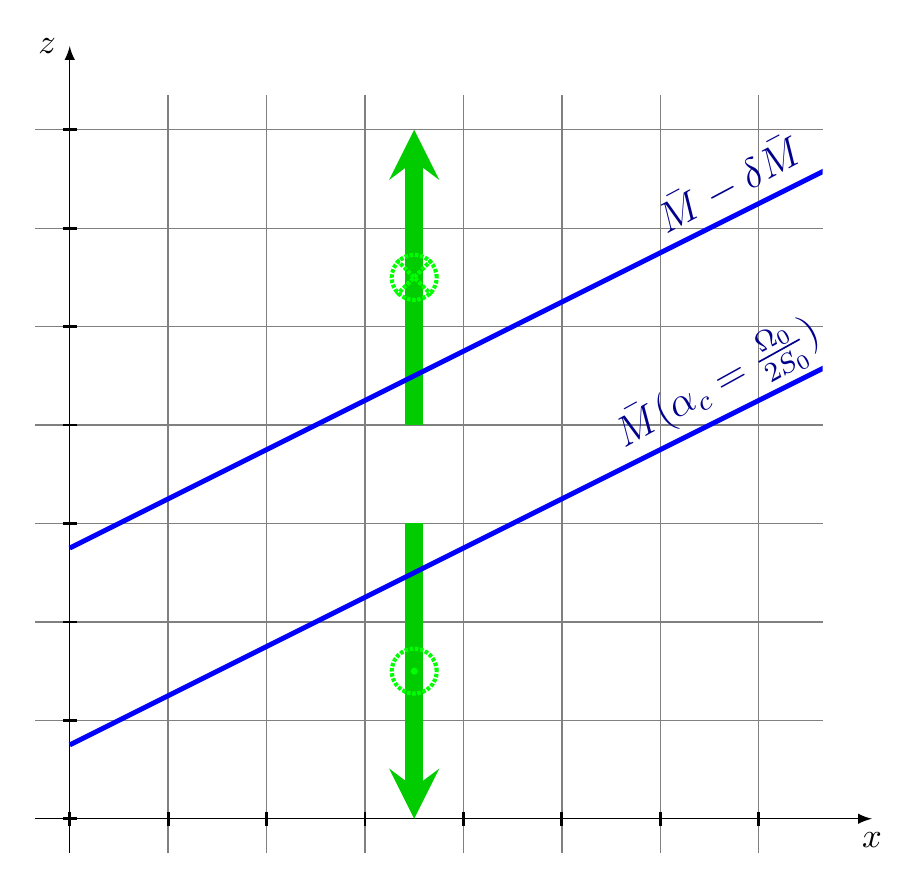}}}%
\subfigure[]{
\resizebox*{8cm}{!}%
{\includegraphics{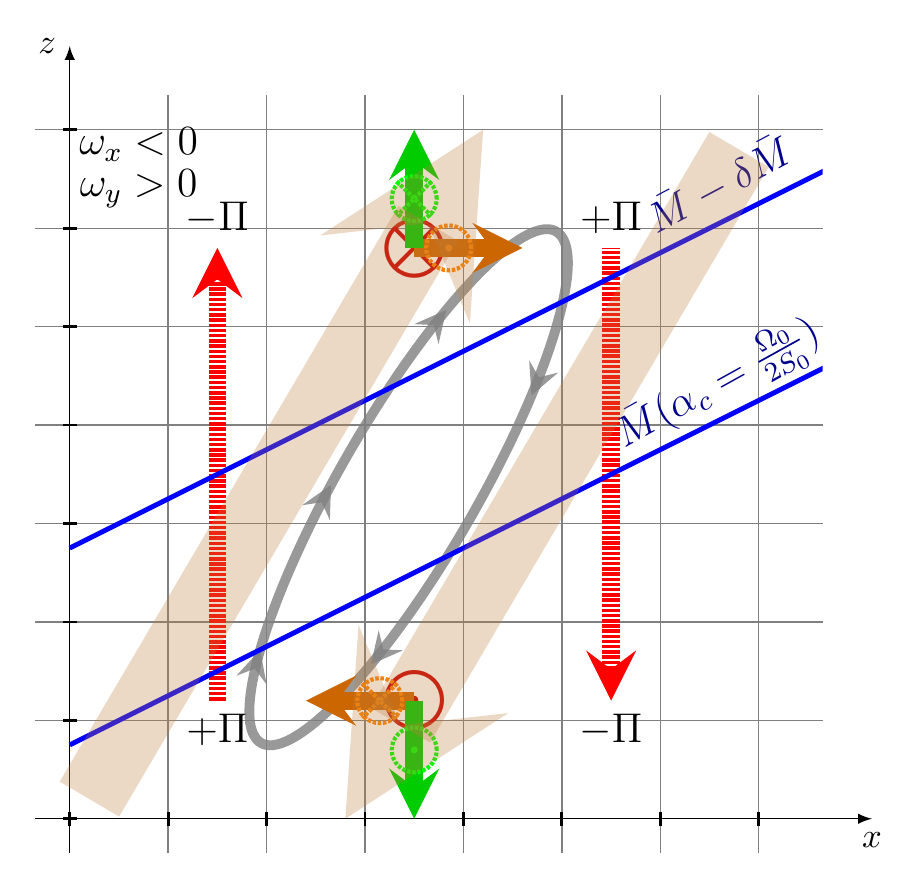}}}%
\caption{VSI instability from vorticity component interaction. (a) For unstable setups $\omega_x$ and $\omega_y$ are in opposite signs. Here negative $\omega_x$ generates positive $\omega_y$ by the mechanism described in figure 3(a). (b) Generation of psotive $\omega_x$ by positive $\omega_y$ via the mechanism described in figure 3(b). (which leads to stability). (c) Generation of negative $\omega_x$ by vertical divergence, via the mechanism described in figure 3(c). (which leads to instability). (d) Superposition of the mechanisms in sub-figures (a-c). When the slantwise motion in the ($x-z$) plane is steeper than the mean absolute momentum slope 
the amplifying mechanism (c) overwhelms the decaying mechanism (b).
In all sub-figures solid line arrows represent the generating mechanisms and dashed or dot arrows represent the responding dynamics. (Colour online)}
\end{center}
\end{figure}

\section{Energy perspective}
The condition for instability ($\alpha >  \alpha_c$)  agrees as well with the Solberg-H{\o}iland Rayleigh like condition \citep{Umurhan_etal_2013,barker2015vertical,latter2018vortices,lin2017thermodynamic}. 

Suppose we exchange two closed-looped\footnote{Because this analysis is axisymmetric, we imagine infinitesimally thin annuli orbiting the central 
star.} fluid filaments located initially at positions 
$(x,z)_1, (x,z)_2$, so that $(\delta x = x_2 - x_1, \delta z = z_2 - z_1)$  and each hoop's slope trajectory is $\alpha = \delta z / \delta x\,$. Denoting the initial state (before the flow is perturbed) and the final state (after the parcels exchanged places) by the superscripts $i,f$, respectively, material conservation of the absolute momentum implies that:  
\be
V^{f}_{1,2} = {\overline V}_{2,1} \pm C_y \delta x.
\ee
The corresponding change in the azimuthal energy of the two parcels due to the exchange is given by,
\be
\Delta E = \dfrac12\[\left.\bigl(V^{f}_{1}\,\bigr)\right.^{\!2} + \left.\bigl(V^{f}_{2}\bigr)\right.^{\!2}\] - \dfrac12\[\({\overline V}_{1}\)^2 + \({\overline V}_{2}\)^2 \]\, .
\ee
Substituting equation (15) and the definitions of $(\alpha, \alpha_c)$ into the above and sorting through the algebra we find
\be
\Delta E = C_y S_0 (\delta x)^2(\alpha_c - \alpha)\, .
\label{DeltaE}
\ee
On the basis of energy minimization,
linear instability is expected when $\Delta E<0$.  Given that both $C_y$ and $S_0$ are defined as positive constants, inspection of equation (\ref{DeltaE}) demonstrates that we recover the previously identified condition for instability, $\alpha > \alpha_c$.
We note that the derivation of this Rayleigh-like condition is somewhat heuristic: As in the Rayleigh conditions for centrifugal or inertial instabilities the hidden assumption is that fluid parcels/filaments instantaneously reach pressure equilibrium. This means that the work performed by the perturbation pressure gradient, which would normally contribute to the energy budget but actually contributes nothing to it because the fluid density is assumed constant, is here ignored.
The consequence of this assumption is that only the azimuthal component of the energy is taken into account as the work performed by the vertical component of the pressure gradient force (hereinafter, PGF) is not considered and the radial velocity is assumed to be tied to the vertical one by continuity. The strength of this rests on the robustness of the assumption of incompressibility, which is more or less valid if
the timescales of fluid motions of a given length scale are much longer than the corresponding sound propagation timescales.

Therefore the Solberg-H{\o}iland condition is equivalent to the condition to obtain growth in the azimuthal component of the perturbation kinetic energy. From (9) we get
\be
\der{}{t}\({v^2 \over 2}\) =  \(1-{\alpha_c \over \alpha}\)S_0 vw\, ,
\label{DisturbanceReynoldsStress}
\ee
indicating that positive Reynolds stress (where $v$ and $w$ are positively correlated in the $(y,z)$ plane, as 
$S_0 = -\upartial{\overline V}/\upartial{z}>0$) leads to energy growth when $\alpha > \alpha_c$.

Since the source of the instability is the vertical mean shear we may define the non-dimensionalized growth rate as $\lambda_{inc} \equiv -{\mathrm i}\omega/S_0$ (where the subscript $inc$ denotes the incompressible solution). Then for the disk values, $C_x = 4C_y$, we obtain from (3) that the growth rate depends solely on the ratio between the fluid parcel's slope trajectory and the mean absolute momentum slope, $\chi \equiv \alpha/\alpha_c$, i.e., 
\be
\lambda_{inc} = {2\over \chi}\sqrt{\chi -1}\, .
\ee
In figure 5 we plot $[\lambda, (u,v,w),(\omega_x,\omega_y)]$ as a function of $\chi$ for the instability regime ($\chi >1$).
The largest growth rate is obtained at $\chi=2$ ($\lambda_{max}(\chi=2) = 1$), that is when the parcel's slope at the $(x,z)$ plane is twice of the critical slope
(${w/u} = 2\alpha_c = \varOmega_0/S_0$). Since $w=2v$ for $\chi=2$, the parcel slope in the $(x,y)$ plane is $\alpha_c$. 
It is evident from figure 5(b,c) that for the most unstable mode the fluid motion in the radial-vertical plane is more prominent than in the azimuthal-vertical plane.

A quick inspection indicates that the largest possible instantaneous growth rate cannot be maintained.
Consider the perturbation kinetic energy growth for the azimuthal and vertical components
\be
\der{}{t}\({v^2 +w^2 \over 2}\) = \(1+ {3\over \chi}\)S_0 vw\, .
\ee
Applying the Cauchy-Schwarz inequality indicates that the largest possible instantaneous normalized growth rate is obtained when $v=w$ with $\lambda_{inst} = [1 +(3/\chi)]\big/2$. 
For the modal instability regime it is maximized to a value of 2 for $\chi = 1$, however this instantaneous growth cannot be sustained since it amplifies $w$ while leaving $v$ unchanged as the right-hand side of (1b) vanishes for $\chi = 1$. Similar non-modal growth dynamics is obtained in atmospheric symmetrically unstable flows \citep{heifetz2008non}. The major difference between these two problems is the lack of explicit dependence in the isentropes' slope in the disk asymptotic equation set (1), and consequently the lack of the condition for mean negative Ertel potential vorticity. The latter is equivalent to the condition for the isentropes' slope to exceed the slope of constant absolute momentum surfaces \citep[e.g.,][]{holton2012introduction}. 
The differences between those two similar problems are discussed in the Appendix where the governing equations of the mesoscale atmospheric symmetric instability (e.g. equation set (1) in \citealt{heifetz2008non}) are translated to the disk shearing box configuration.  

\begin{figure}[ht!]
\begin{center}
\subfigure[]{
\resizebox*{8cm}{!}%
{\includegraphics{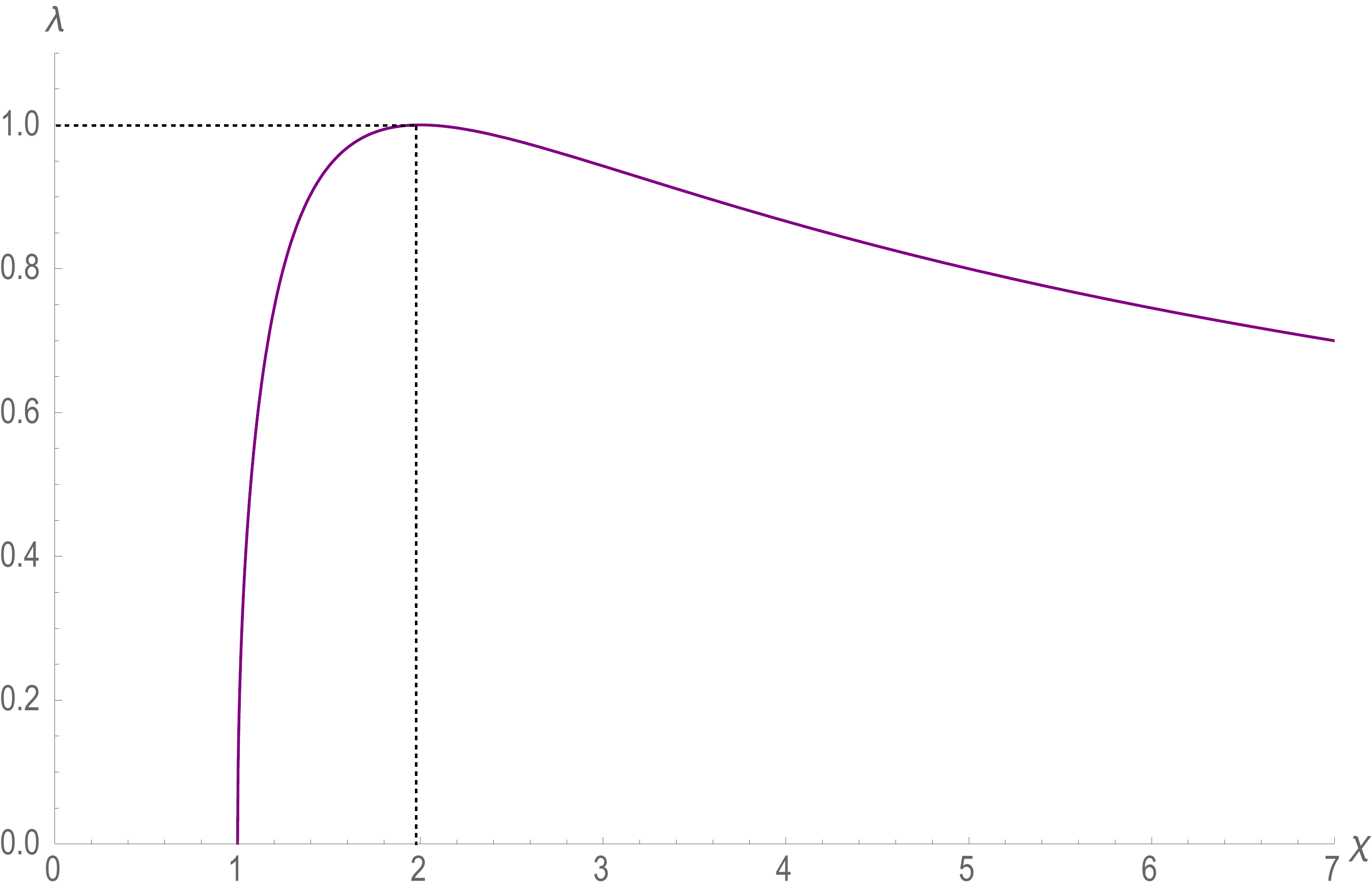}}}%
\subfigure[]{
\resizebox*{8cm}{!}%
{\includegraphics{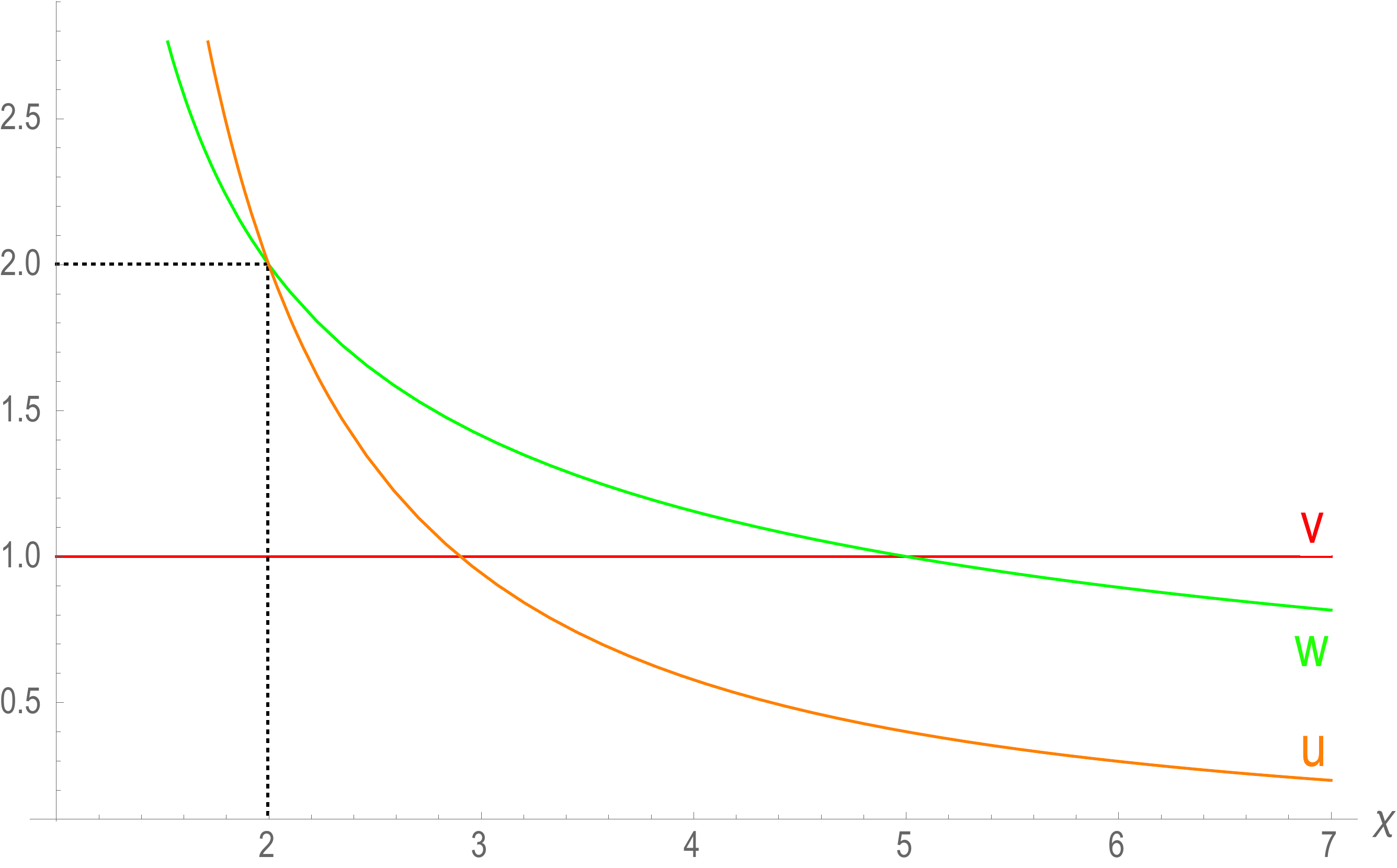}}}%
\vspace{0.5cm}
\subfigure[]{
\resizebox*{8cm}{!}%
{\includegraphics{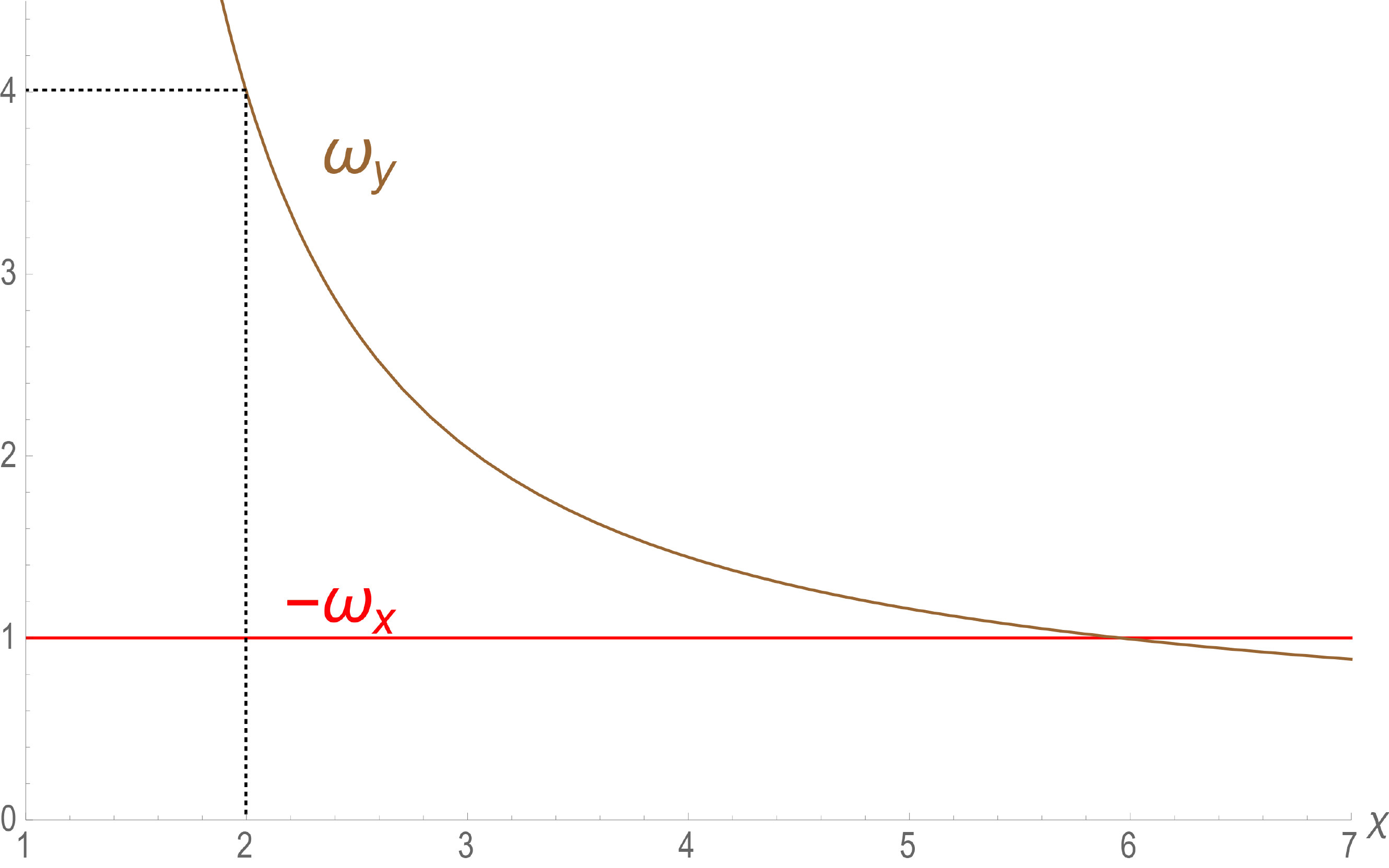}}}%
\caption{(a). Modal growth rate $\lambda$, as a function of $\chi \equiv \alpha/\alpha_c$, equation (19). (b) and (c). Velocity and vorticity components of the corresponding normal mode solutions. (Colour online)}
\end{center}
\end{figure}

\section{Oscillatory instability resulting from anelasticity}

Detailed numerical simulations of protoplanetary disk ``dead zones'' show the existence of oscillatory unstable VSI dynamics
\citep{2014A&A...572A..77S,barker2015vertical,2016MNRAS.456.3571R,stoll2016particle} suggested to be rooted in 
anelasticity \citep{Umurhan_etal_2016}. 
Typically speaking, overstable dynamics cannot be described by the incompressible analysis like what we have employed up to now. 
Indeed, equations (3) and (19), based purely on an assumption of incompressibly, indicate that the modes can be either exponentially growing or purely oscillatory, but never as oscillations with exponentially growing amplitude (overstability). 
Here we show that if we replace incompressibility with anelasticity, we recover overstable behaviour. 

An analytically tractable solution is obtained by assuming a neutrally-stratified exponential density profile in the form of ${\overline \rho} = \rho_{m}\exp(-z/H)$, where $\rho_{m}$ is the mid-plane density and the density scale height, $H$, is assumed constant. Substituting this form for $\overline\rho$ into equation (1d) we obtain
\be
\der{u}{x}+ \der{w}{z} - {w\over H} = 0\, ,
\ee
where consequently, equation (2) has now an additional term in the right-hand side
\be
\der{^2 \ddot \varPi }{x^2} = -C_x\[S_0{\upartial {^2 \varPi }\over \upartial {x} \upartial {z}} + C_y \(\der{^2 \varPi }{z^2}  -{1\over H}\der{\varPi }{z}\)\].
\ee
In order to properly account for stratification in the perturbation energy we introduce the modified plane wave solution of the form of 
$\exp[{\mathrm i}(kx +mz -\omega t)+z/2H]$. Inserting into equation (22) we obtain the complex dispersion relation
\be
\omega^2 = {S_0 C_x \over \alpha ^2}\biggl\{  \alpha_c \left[1+\left(\frac{\alpha}{2kH}\right)^{\!2}\right]- \alpha\left[1+{\mathrm i}\,\frac{\alpha }{2kH}\right] \biggr\}\,.
\ee
The equivalent VSI condition to equation (4) for the ``exponential atmosphere'' is derived in Appendix B
\be
\left | \frac{w}{u}\right |=\frac{\alpha}{\sqrt{1+\beta^2}}>\alpha_c\sqrt{1+\beta^2}\hskip 8mm \Longrightarrow\hskip8mm \operatorname{Re}\left\{\frac{w}{u}\right\}>\alpha_c
\ee
(see particularly (B.1), (B.8) and (B.9)), where $\alpha\equiv-k/m$, and $\beta =1/(2mH)$. For infinite scale height $H$, equation (4) is recovered. For finite $H$ however, the mean parcel's slope is smaller than the ratio $k/m$ and by the same time the minimal critical slope for instability increases. Therefore the anelastic effect tends to inhibit VSI.

Using the disk values of $C_x = 4C_y$  we obtain a complex normalized growth rate wherein $\lambda_{ane} \equiv \lambda_r +{\mathrm i}\lambda_i = -{\mathrm  i}\omega/S_0$ (where the `ane' subscript refers to the anelastic solution), in which $\lambda_{ane}$ satisfies the following
\be
\lambda^2_{ane} = \[\lambda^2_{inc} - \({\alpha_c \over kH}\)^2\] +2{\mathrm i}\({\alpha_c \over kH}\) = \(\lambda^2_r - \lambda^2_i\) + 2{\mathrm i}\lambda_r\lambda_i.
\ee
Equation (25) admits solutions in which perturbations may grow while oscillating (overstability). Note that for positive growth rate, 
$\lambda_r >0$, the imaginary part of (25) indicates that $\lambda_i >0$ as well. Normalizing time by $S^{-1}_0$ the general structure of a growing modal perturbation is of the form 
$\exp(|\lambda_r|t) \exp(z/2H)\exp[{\mathrm i}(kx +mz +|\lambda_i| t)]$.
A straightforward derivation (not shown here) indicates that $\lambda_r<1$ for finite scale height, thus, in agreement with (24) and with \citet{mcnally2015vertically}, the anelastic effect tends to decrease the instability. We focus now on the mechanism by which anelasticity generates oscillations. To isolate the latter we can look at the untilted limit
where $m=0$ so that $\lambda_{inc}=0$. Then the oscillation is due to propagation in the radial direction with the normalized phase speed $c_x = - |\lambda_i|/k$.    
A further simplification of the algebra is obtained when considering the wavenumber $k$, satisfying ${\alpha_c /(kH)}=1$.
Then  $\lambda^2_{ane} = -1 +2{\mathrm i}$, with $(\lambda_r, \lambda_i) = (0.786, 1.27)$.
The continuity equation (21) dictates that the divergence in the $(x,z)$ plane is proportional to $w$ and for $m=0$, $u=-{\mathrm i}w/(2kH)$. 
Furthermore writing $(v,w)$ in terms of their amplitude and phases
\be
v= \hat{v}(t)\exp\{{\mathrm i}[kx +\epsilon_v(t)]\}\,, \hspace{1.0cm} w= \hat{w}(t)\exp\{{\mathrm i}[kx +\epsilon_w(t)]\}\, ,
\ee
the eigen-structure satisfies $\hat{v} \approx 0.75\hat{w}\,$; $(\epsilon_w -\epsilon_v) \approx 0.176\pi$. By geostrophy 
$v$ lags the pressure perturbation $\varPi $ by a quarter of a wavelength. This normal mode structure is plotted schematically in figure 6(a). 

We wish to explain now how this mode is growing while propagating in concert.
Substitute it into equation (1b) we obtain   
\be
\dot{v} = \bigl(\tfrac12{\mathrm i}+1\bigr)
w \approx 1.12 \,{\mathrm e}^{0.15\pi {\mathrm i}} w,
\ee
indicating that the generation of $v$ is to the left of $w$ (figure 6(b)) since both positive $w$ and negative $u$ generate positive 
$v$, and negative values of  $u$ are located a quarter of wavelength to the left of positive values of $w$ (figure 6(a)).
On the other hand, using the gesotrophy conditions of equation (1a) and equation (1c) we find that
\be
\dot{w} = 2\, {\mathrm e}^{0.5\pi {\mathrm i}}v,
\ee
i.e., the vertical pressure gradient anomaly associated with the geostrophic balance of $v$, shifts $w$ a quarter of wavelength to the left of $v$ (figure 6(c)). As a result both $v$ and $w$ are growing while propagating to the left. Put more rigorously, we can substitute $v$ and $w$ into equations (27) and (28). Then we solve for the real and the imaginary parts. 
This gives for modal instability (where all fields experience the same growth rate and propagation speed)
\be
\lambda_r = {\dot{\hat{v}} \over \hat{v}} = {\dot{\hat{w}} \over \hat{w}} = 
2{\hat{v} \over \hat{w}}\sin{(\epsilon_w -\epsilon_v)}\, ,  \hspace{1.0cm} 
\lambda_i = \dot{\epsilon}_v  = \dot{\epsilon}_w = 2{\hat{v} \over \hat{w}}\cos{(\epsilon_w -\epsilon_v)}\, .
\ee
For completeness we note that in this untilted solution $\omega_y = [1+(2kH)^2]u/(2H)$ and $\omega_x = -v/ (2H)$. Hence the normalized vorticity equations become  
\be
{\dot \omega}_y = -\[{1+ (2\alpha_c)^{2} \over \alpha_c }\]\omega_x,
\ee
and
\be
{\dot \omega}_x =  - (2{\mathrm i}-1)\[{1+ (2\alpha_c)^{2} \over \alpha_c }\]^{-1} \omega_y .
\ee
Similar to the incompressible case, when $\omega_x$ and $\omega_y$ are anti-phased the two vorticity components amplify each other, but in the anelastic case they also shift each other in the $x$ direction to propagate in concert. 
\begin{figure}[ht!]
\subfigure[]{
\resizebox*{8cm}{!}%
{\includegraphics{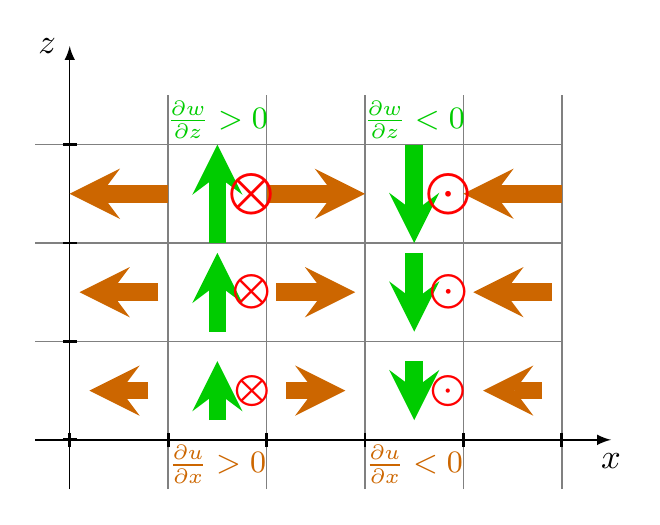}}}%
\subfigure[]{
\resizebox*{8cm}{!}%
{\includegraphics{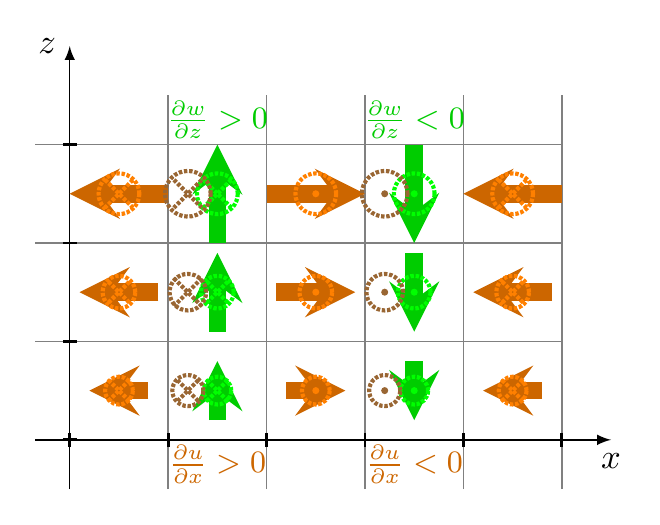}}}%
\vspace{0.5cm}
\subfigure[]{
\resizebox*{8cm}{!}%
{\includegraphics{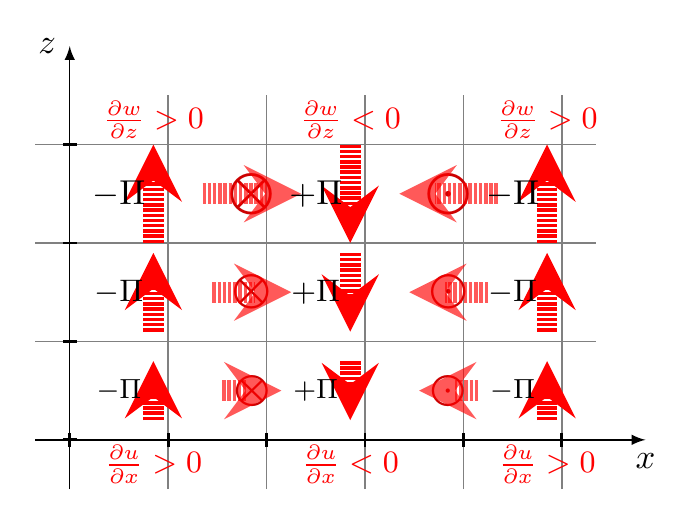}}}%
\subfigure[]{
\resizebox*{8cm}{!}%
{\includegraphics{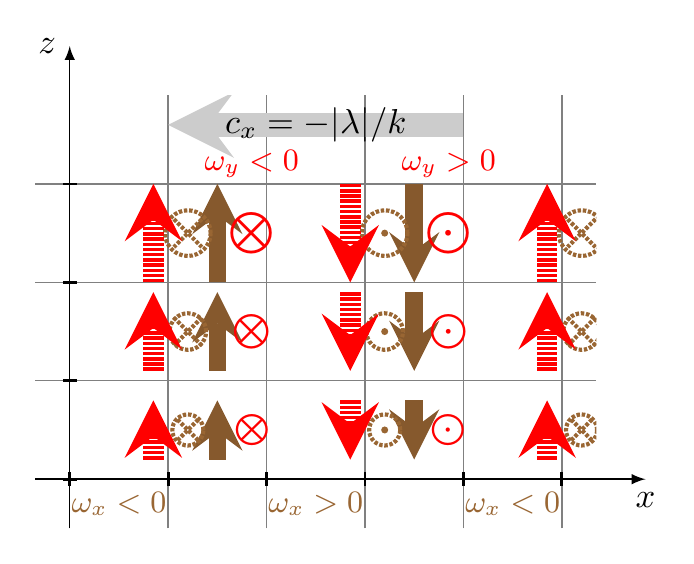}}}%
\caption{Propagation and growth mechanisms of untilted ($m=0$), anelastic modes. (a). Structure of the modal velocity components presented in the $(x,z)$ plane. (b). Generation of azimuthal velocity due to the combined effects of the vertical advection of mean azimuthal velocity and the Coriolis force acting on the radial velocity.  (c). Generation of vertical velocity due to the pressure structure imposed by geostrophy  and the radial velocity which is tied to it via anelastic continuity. (d). Inward propagation combined with growth, due to the combined effects described in subplots (a-c). For clarity only the azimuthal and vertical velocities are plotted. 
In all sub-figures solid line arrows represent the generating mechanisms and dashed or dot arrows represent the responding dynamics. (Colour online)}
\end{figure}

\section{VSI in a realistic disk setup}

We further investigate the VSI in a more relevant setting for a planet-forming disk.  We therefore assume a locally isothermal atmosphere (i.e., $T$ is a function of radius only) with a Gaussian density profile in the form of $\bar{\rho}=\rho_m\exp(-z^{2} /2 H^2)$ \citep{2013MNRAS.435.2610N,Umurhan_etal_2016}. Substituting $\bar\rho$ into equation set (1) results in the following equation set in dimensional form:
\bse
\begin{align}
0 =\,&\,2 \varOmega v-\frac{\upartial {\varPi }}{\upartial x} ,
\label{diskxmom}   \\
\frac{\upartial v}{\upartial t} =\,&\,-\frac{1}{2}\varOmega  {u} -\frac{1}{2}q\varOmega\epsilon \dfrac{z}{H}w ,
\label{diskymom}  \\
  \frac{\upartial w}{\upartial t} =\,&\,-\frac{\upartial {\varPi }}{\upartial z} ,
  \label{diskzmom}  \\
0 =\,&\,\frac{\upartial u}{\upartial x}+\frac{\upartial w}{\upartial z}-\dfrac{z}{H^2} w \,,
\label{diskcontinuity}
\end{align}
\ese
where for ${\overline S}(z)$ we have used $-{\upartial\overline{v}}/{\upartial z}=-{qz\varOmega}\epsilon/{2H}$.
This set is properly non-dimensionalized by assuming the following scalings for space and time as $x \rightarrow \epsilon H \tilde x, \ z \rightarrow H \tilde z, \ t \rightarrow \varOmega\tau/\epsilon$,
while for the other depended quantities we assume $u \rightarrow \epsilon \tilde u, \ 
w \rightarrow \tilde w, \ v \rightarrow \tilde v$ and $\varPi  \rightarrow \epsilon \tilde \varPi $, in which all quantities with tildes over them are non-dimensional.  We then rewrite equations in terms of those quantities and we drop all tildes except for $\tilde u$ and $\tilde \varPi $ (in order to visually track that these dependent quantities are necessarily small by a factor of $\epsilon$ in comparison to the other quantities).\par
After dropping the tilde symbol from all independent variables so that $t,x,z$ are in terms of the above introduced scalings, we combine equations (\ref{diskxmom}--d)
into a single partial differential equation governing the response of $\varPi $, i.e.,
\renewcommand{\theequation}{\arabic{equation}}
\begin{equation}
    -\frac{\upartial^{2}}{\upartial t^{2}} \biggl(\frac{\upartial^{2} {\varPi }}{\upartial x^{2}}\biggr)-\frac{\upartial^{2} {\varPi }}{\upartial z^{2}}+\left(1+q \frac{\upartial}{\upartial x}\right) z \frac{\upartial {\varPi }}{\upartial z}=0 \, .
    \label{tilde_Pi_eqn}
\end{equation}
We assume the usual normal mode ansatz
\be
\varPi =\hat\varPi (x,z){\mathrm e}^{-{\mathrm i} \omega t} + {\mathrm{c.c.}} \,,
\label{tilde_Pi_ansatz}
\ee
where we consider solutions for $\hat\varPi $ that both filters out surface modes and retains the body modes  
by adopting the the following ansatz
\be
\hat\varPi =P_m(z,x)=
\displaystyle
\left\{
\begin{array}{cc}
\sum_{n=0,2,\bm\cdots}^{m} P_{n,m}(x)z^n, &\quad m = {\rm even}, \\[0.4em]
\sum_{n=1,3,\bm\cdots}^{m} P_{n,m}(x)z^n, &\quad m = {\rm odd},\
\end{array}
\right.
\label{hat_Pi_solution}
\ee
\citep[as in][]{Umurhan_etal_2016}\footnote{
Note, equation (5) of \citet{Umurhan_etal_2016} contains a typographical error: the sign in front of the second partial derivative with respect to $z$ term should be negative, cf. equation (\ref{tilde_Pi_eqn}) of this study.  All results quoted in that paper, especially those pertaining to the eigenvalues calculated, are unaffected the sign of this term as it does not enter into its determination.}. 
As most of the 
kinetic energy of the instability is contained in the low order body modes, we only consider  modes with  $n=m=1,2$ (to avoid confusion please note that in this section $m$ is not the vertical wavenumber), subject to the no normal flow boundary condition at the inner and outer edges, which in terms of this formulation becomes
\be
\frac{\upartial P_{n, m}}{\upartial x}+\frac{qn}{\omega^{2}} P_{n, m}=0 \qquad\left(\text{at } x=\pm L_{x}\right),
\ee
in which
\be
\frac{\omega^{2}}{m}=\frac{1 \pm \sqrt{1-k_{j}^{2}q^2}}{2 k_{j}^{2}} \,.
\label{growth_rate_disk}
\ee
The boundary condition in equation (36) dictates that this model of the VSI is vertically global but radially local and therefore only appropriate for studying narrow radial disk patches. While a more appropriate boundary condition might be to ignore radial boundaries and seek (suitably slanted) locally wave-like perturbations, we impose the no normal flow boundary condition as these are in the same spirit as those of published simulations. In addition, the choice of the boundary condition is important for the details of the manifestation of the instability, but it does not affect its underlying mechanism, for further justification \citep[see][]{Umurhan_etal_2016}.
For $m=1$ the function sum $P_1$ consists of the single
separable function $P_{1,1}z$, while for $m=2$
 $P_2$ is given by the inseparable form $P_{2,2}z^2 + P_{0,2}$.  In 
 general the ``top" functional form for any value $m$ is given by
\be
P_{m,m}=\Big[A_j\sin(k_jx)+B_j\cos(k_jx)\Big] \exp{\displaystyle\left(-\frac{qm}{2\omega^2}x\right)} \,.
\label{Pm_solution}
\ee
where $k=k_{j} \equiv j \pi / 2 L_{x},$ where $j$ is any integer including zero. When $j$ is an odd integer, then $A_{j}=-k_{j}, \quad B_{j}=$ $ -qm / 2 \omega^{2} .$ However, when $j$ is an even integer (including zero) $A_{j}= -qm / 2 \omega^{2}, \quad B_{j}=-k_{j}$.\par
 
For $P_2$  the additional term $P_{0,2}$ is calculated by solving
\be
\omega^{2} \frac{\upartial^{2} P_{n, m}}{\upartial x^{2}}+n\left(P_{n, m}+q \frac{\upartial P_{n, m}}{\upartial x}\right)= -(n+2)(n+1) P_{n+2, m}
\ee
subject to the boundary condition (36). It results in
\be
P_{0,2}=\tfrac{1}{2} (-1)^{2/3} \left(\sqrt{3}+{\mathrm i}\right) \exp\bigl(2 \sqrt[3]{-1}\, x\bigr)
   \left[\left(\sqrt{3}+{\mathrm i}\right) \sin (2 x)+2 {\mathrm i} \cos (2 x)\right]
\ee
for $k=2$.
\par

 We note that only the $m=1$ fundamental corrugation mode is a separable functions of $x$ and $z$. However, this
 feature does not carry over to the structure functions for all overtones (i.e., $P_m(z,x) \ \forall \ m \ge 2$) -- these will always be functions of $x$ and $z$ that are not fundamentally separable in the usual sense of of its definition.  However we note that given that these are a polynomial series in 
 $z$, these functions are easily factorizable.\par
\par 
To calculate the ratio ${w}\big/{u}$ we explicitly use equations (\ref{diskxmom}--c) and 
(\ref{Pm_solution}),
to produce
\begin{align}
  \frac{w}{u}\, = \,\frac{\tilde w}{\epsilon \tilde u} =\,&\, \frac{1}{2 \epsilon}\left[\frac{ \upartial_{z} \hat{\varPi } / {\mathrm i} \omega+{\mathrm{c.c.}}}{\left[\left({\mathrm i} \omega / 2 \right) \upartial_{x} \hat{\varPi }-\left(1 / 2 {\mathrm i} \omega\right)  qz \upartial_{z} \hat{\varPi }\right]+{\mathrm{c.c.}}}\right] \nonumber\\
= \,&\, \frac{1}{ \epsilon} \left[\frac{ \upartial_{z} \hat{\varPi }+{\mathrm{c.c.}}}{\left[-\omega^{2} \upartial_{x} \hat{\varPi }-q  z \upartial_{z} \hat{\varPi }\right]+{\mathrm{c.c.}}}\right]  .
\label{rat_w_over_u}
\end{align}
Inserting the expression for $\hat\varPi $, found 
in equations
(\ref{tilde_Pi_ansatz}-\ref{hat_Pi_solution}),
into equation (\ref{rat_w_over_u}) reveals that ${w}\big/{u}$ is dependent on $m$ 
because of this inseparability of the individual eigenfunctions 
describing $u$ and $w$.
While overtone modes (i.e., $\forall m>2$) are of academic interest, our concern here is just with the fundamental modes as these are the ones that appear prominent in disk simulations of the VSI \citep[e.g.,][and others]{2013MNRAS.435.2610N,2016MNRAS.456.3571R,manger2018vortex}. 
\par
In figure (7) we plot the inverse tangent of the radial average of $w/u$
-- here as $\big<w\big/u\big>$ --
for $P_1$ and $P_2$, together with the inverse tangent of $\alpha_c$, as functions of $z$ for the maximum growth rate wavenumber
\makeatletter
\newcommand{\xRightarrow}[2][]{\ext@arrow 0359\Rightarrowfill@{#1}{#2}}
\makeatother
 \begin{equation}
    \left|k_{\max }\right|=\frac{2}{|q|} \hskip8mm \xRightarrow[]{\bigl.q=-1\bigr.} \hskip8mm k_{max}=2 
\end{equation}
(in dimensional units the fastest growing modes
scale as $\lambda_{{\rm max}} = \pi \epsilon H$).
As indicated from figure (7a), between 
$0<z<1$, the corrugation mode is potentially unstable while the breathing mode is predicted to be stable. Where
$z>1$ we observe $\pi/2>\big<w\big/u\big>>\alpha_c$ for both the corrugation and the breathing modes, suggesting that this region supports unstable dynamics for the two modes. $\big<w\big/u\big>$ of the two modes decrease with height and converge at $z\approx 5$ to the approximate angle of $\pi/5$, where the threshold stability slope, $\alpha_c = 1/z$, decreases with height as well.
Consequently, one might speculate that the VSI should initially manifest itself at $1<z<2$ by the breathing mode -- whose growth rate is also a bit larger than the corrugation mode, as seen from equation (\ref{growth_rate_disk}) -- while the corrugation mode can spread/initiate the instability in the lower regions of the disk. This seems to be consistent with simulations showing the VSI initially taking shape around 1 scale height via a breathing mode and then evolves into a corrugation mode which spread into regions above and below \citep[]{2013MNRAS.435.2610N,2014A&A...572A..77S,2019PASP..131g2001L}. figure 7(b) is the same as figure (7a) but plotted up to $z=100$. While typical disks in the Ohmic Zone do not span further than a few scale heights it can be seen how at high values of $z$ the slopes converge to the stable threshold of $\alpha_c$.

\begin{figure}[ht!]
\begin{center}
\subfigure[]{
\resizebox*{14cm}{!}%
{\includegraphics{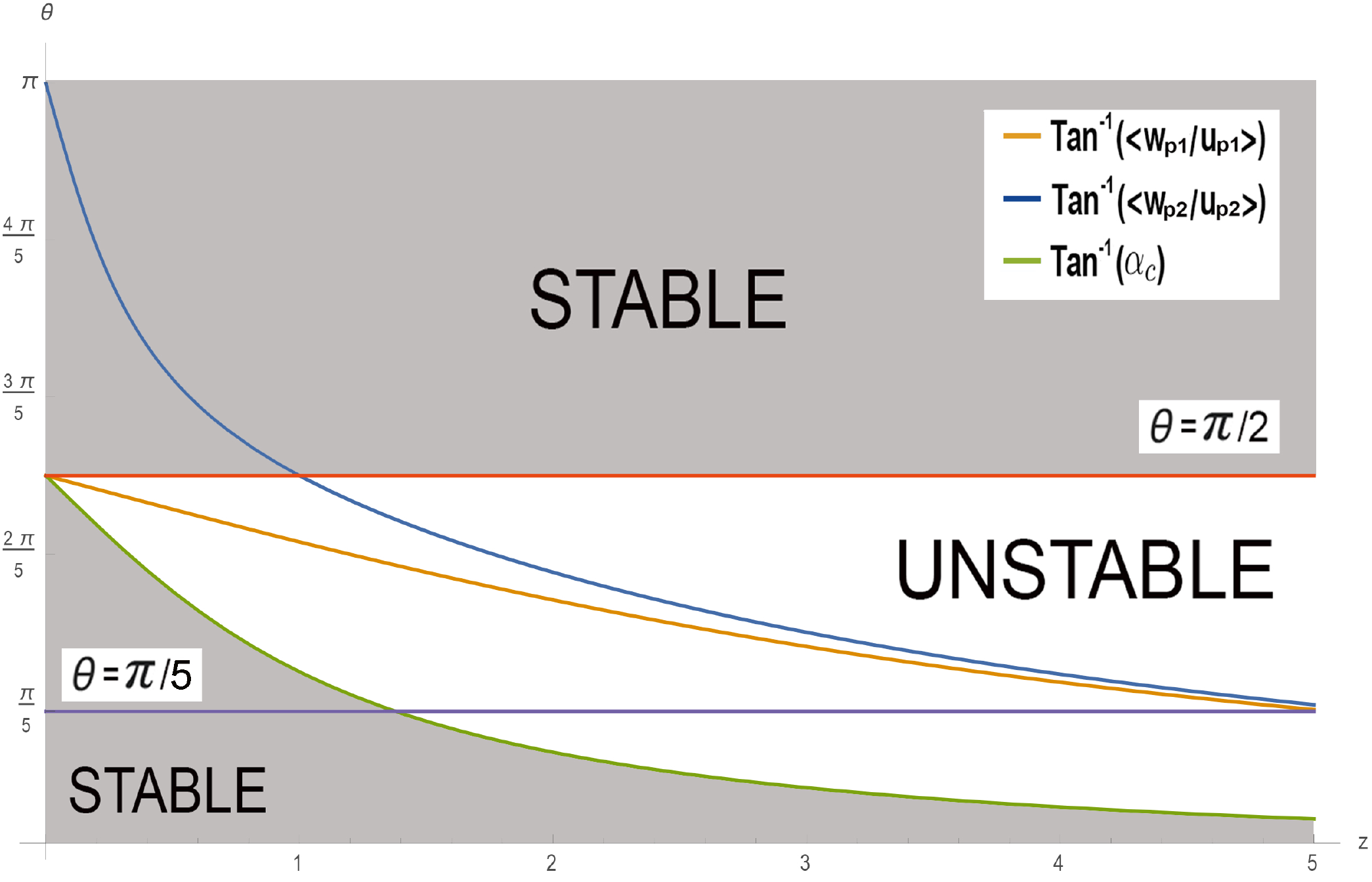}}}%
\\
\subfigure[]{
\resizebox*{14cm}{!}%
{\includegraphics{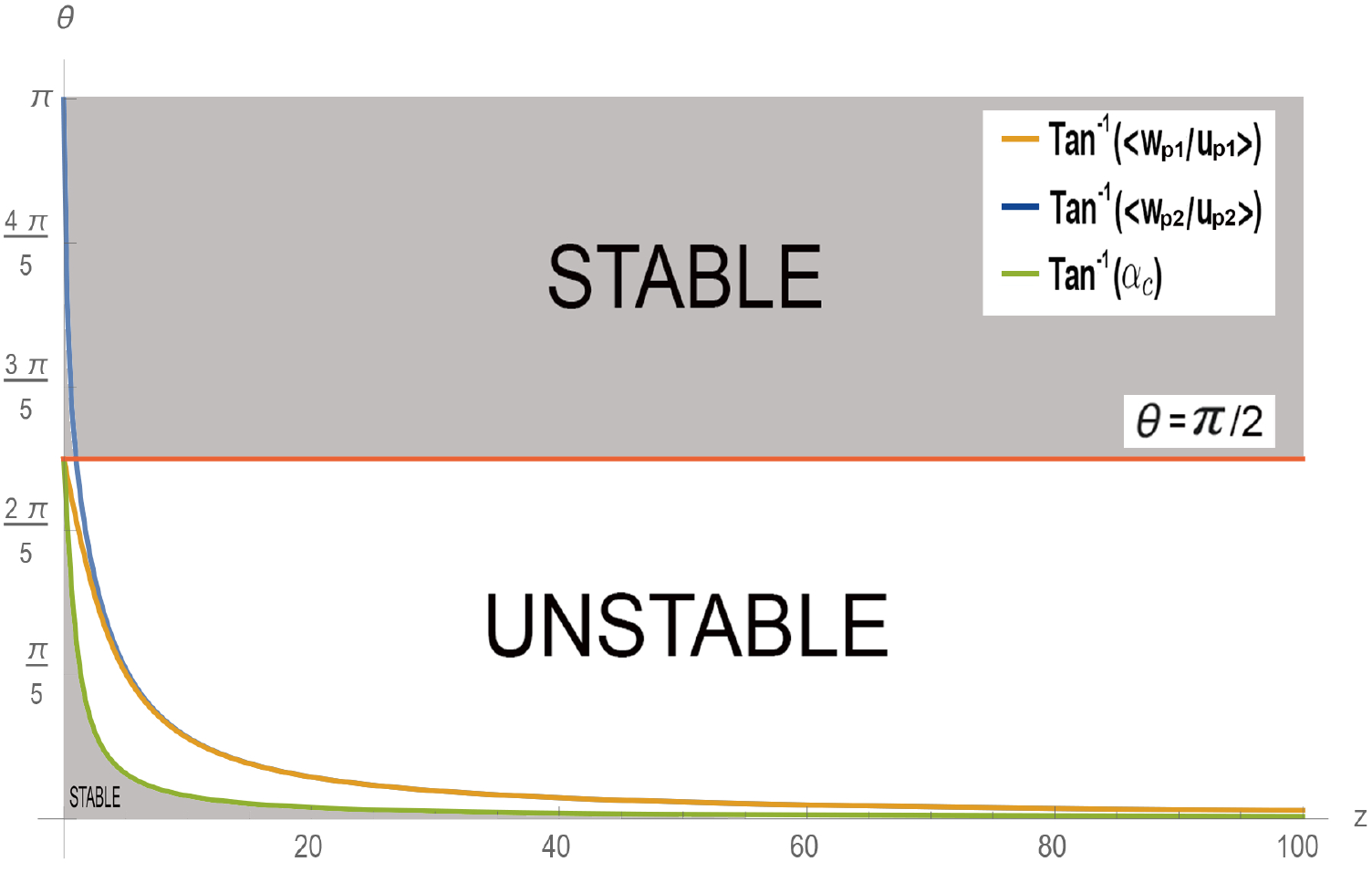}}}%
\caption{The R.H.S of equation (41) plotted for the corrugation $P_1$(orange), and breathing $P_2$(blue) modes and the inverse tangent of the critical slope $\alpha_{c}=1/z$ (green), all as a function of $z$. (a, top panel) $0<z<5$ (b, bottom panel) $0<z<100$, the plots are in the z-$\theta$ surface where $q=-1$ and $k=k_{max}=2$, for the setup of Gaussian density profile. (Colour online)}
\end{center}
\end{figure}

\section{Summary and conclusions}
Detailed numerical simulations of the dead zone dynamics of cold protoplanetary accretion disks have demonstrated that VSI plays an essential role in driving turbulent gas motions that can stir solids and affect the planet formation process.
As the simulations are highly complex, with several processes occurring during various stages of their development, identifying the underlying physical mechanisms 
responsible for the wide phenomena manifest in them
are not always clear. This is especially valid for the linear instability itself.
For this reason we have devised a physically meaningful minimal model of the VSI that captures the essence of the instability mechanism while remaining analytically tractable.  

Since 2D (barotropic) Keplerian shear flows are stable for small perturbations, the VSI can be regarded as a mechanism that overcomes this obstacle by exploiting the vertical shear within the disk that emerges as a baroclinic response to a radial temperature gradient. With respect to the disk's radial-vertical planes a fluid parcel's motion should be particularly slantwised:  On one hand, it should be high enough to obtain more energy from the vertical shear than it loses to the Keplerian shear by radial motions. On the other hand, it should not be too high for otherwise radial motions are suppressed which is important to foster radial redistribution of the azimuthal angular momentum and maximize energetic growth during the accretion process -- cf.
equation (\ref{DisturbanceReynoldsStress}). Therefore, parcel motions ought be at an angle which is larger than the critical slope while not being too steep for otherwise the rate of energy extraction starts becoming inefficient. In this minimal model the critical slope is determined directly by the ratio between the Keplerian and the vertical shear magnitudes and is also aligned with the surfaces of the mean absolute momentum (angular momentum in a frame of rest) surfaces. 

Here we have shown explicitly how motions exceeding the critical slope accelerates the flow away from its initial position, not only in the radial-vertical plane but also in the azimuthal direction.
Furthermore we have shown that the VSI can be explained in terms of the mutual amplification mechanism between
the 
radial-vertical plane oriented and the azimuthal-vertical plane oriented circulations (quantified by 
their respective vorticity components).
We have also made concrete the energetic perspective of VSI, commonly rationalized in terms of fulfilling the semi-quantitative condition of spontaneous energy release when pairs of fluid parcels exchange positions. This argument is based (like all the generalised Rayleigh-like instability conditions) on instantaneous pressure adjustment of the fluid parcel and consequently on vanishing of the pressure perturbation. This is not the case however for VSI, as is evident from the azimuthal and vertical momentum equations (1a,c). We showed that the Solberg-H{\o}iland condition  is indeed valid but it is the Reynolds stress mechanism, in the azimuthal-vertical plane, that yanks kinetic energy from the vertical mean shear to the perturbation. Furthermore, by the same time the Reynolds stress in the azimuthal-radial plane acts to stabilize the perturbations by transferring energy from the perturbation back to the Keplerian flow. Hence, only when the parcel slope is larger than the critical slope the Reynolds stress mechanism yields instability.
Despite of its simplicity the minimal model enables super-modal transient growth, that is a growth mechanism which is more efficient than the one obtained by the most unstable normal mode. Here we showed how this mechanism operates but also why it cannot be sustained. 

The minimal model in its simplest incompressible guise cannot describe the oscillatory instability dynamics observed in disk numerical simulations. However, we show that it is enough to add neutrally stable anelastic stratification to enable oscillatory instability. The simplicity of the model allows one then to disentangle the oscillatory mechanism, due to the stratification, from the instability one when considering untilted modal structures.

 It is interesting to note the parallels between the VSI and the mesoscale Symmetric Instability slant-wise convection, the latter of which operates in the mid-latitudinal Earth atmosphere as well as being responsible for mixing in the Gulf Stream \citep[e.g.,][]{Thomas_etal_2013}. In Appendix A we converted the equations from its atmospheric context into the disk shearing box setup. The similarity between the equations of the two instability mechanisms becomes transparent as well as the main difference between the two - in Symmetric Instability buoyancy dynamics is explicit, thus consequently, the most unstable mode parcels’ slope coincides with the mean isentropes’ slope. In contrast, in VSI, the buoyancy force is smaller by at least one order of magnitude than the vertical component of the vertical component of the pressure gradient force perturbation. Consequently, the buoyancy force does not play an effective role as a restoring force to inhibit the instability mechanism.    
For future work it could be interesting to generalise the VSI minimal model by assuming the mean flow vertical shear to be in a thermal wind balance that is maintained by the slanted mean isentropes. In such a case the isentropes may be tilted in a different slope from the absolute momentum surfaces. This setup is expected to invoke richer potential vorticity-like dynamical behavior, though its relevancy to the Ohmic Zone should be carefully examined.   

It remains a particularly outstanding
question how dust loading influences the progress of any given  primary
instability mechanism that leads to turbulence.  Two recent studies \citep{Lin_2019,Schafer_etal_2020} examined the role that particle-loading has on 
the emergence and development of the VSI in model disks. The numerical simulations of \citet{Lin_2019} suggest that particles can reduce the efficacy of the VSI.  
The corresponding simulations of \citet{Schafer_etal_2020}, however, indicate that the
particle accumulation mechanism known as the streaming instability \citep{Youdin_Goodman_2005,Johansen_etal_2007} can operate in the
face of the unsteady motions emerging from the VSI, in contradiction to several 
theoretical predictions \citep[e.g.,][]{Umurhan_etal_2020,Chen_Lin_2020,Gole_etal_2020}. 
This controversy requires properly understanding how dust interacts with the fluid motions that gives rise to the VSI and can be a subject for future studies. 
\par
\bigskip
\bibliographystyle{gGAF}
\bibliography{GGAF-2020-0018_Yellin-bergovoy_ref.bib}{}
\markboth{\rm R. YELLIN-BERGOVOY ET AL}{\rm GEOPHYSICAL \&  ASTROPHYSICAL FLUID DYNAMICS}

\appendix
\section{VSI VERSUS ATMOSPHERIC SYMMETRIC INSTABILITY}

Here we translate the governing equations of mesoscale symmetric instability (e.g., equation set (1) in Heifetz and Farrell 2008) to the disk shearing box setup in order to highlight the similarities and differences between them and equation set (1) in this paper. 

First we note that the mean flow in both cases is in thermal wind balance (i.e. geostrophic in the radial direction and hydrostatic in the vertical one)
\be
{\overline V} = {1\over C_x}\der{\overline \varPi }{x}\, , \hspace{1.0cm}  {\overline B} = \der{\overline \varPi }{z}
\hspace{1.0cm} \Longrightarrow \hspace{1.0cm} S = - \der{\overline V}{z} = - {1\over C_x}\der{\overline B}{x}\, ,
\ee
where $B$ is the mean buoyancy. Then the symmetric instability equations in the shearing box become
\begin{align}
\dot{u} = \,&\,C_x v - \der{\varPi }{x},\\
\dot{v} = \,&\,-C_y u + S w,\\
0 =\,&\, - \der{\varPi }{z} +b,\\
\dot{b} =\,&\, -N^2 w  -\der{\overline B}{x}u = -N^2 w  +C_x S u ,\\
\der{u}{x} + \der{w}{z} =\,&\, 0,
\end{align}
where $N$ is the buoyancy frequency and $b$ is the buoyancy perturbation.
Comparing the incompressible version of equation set (1) to (A2)--(A6) we find that in the atmospheric symmetric instability $v$ is not assumed to be in geostrophic balance but instead $w$ is in quasi-hydrostatic balance. The buoyancy dynamics becomes explicit and consequently, for the most unstable mode, the parcels' slope coincides with the mean istentorpes' slope.

 \section{Derivation of the VSI criteria for exponential disk atmosphere}
Substitute the modal solution form $\exp[{\mathrm i}(kx +mz -\omega t)+z/2H]$ in equation (21) for the exponential density profile yields
\begin{equation}
   \frac{w}{u}=\frac{\alpha}{1+\beta^2}(1-{\mathrm i}\beta)=\frac{\alpha}{\sqrt{1+\beta^2}}\tan^{-1}(-\beta) ,
\end{equation}
where $\beta=1/(2mH)=-\alpha/(2kH)$ and $\alpha=-k/m$. Rewrite the dispersion relation of equation (23) as
\begin{equation}
\omega^2 = D\bigl[  \alpha_c(1+\beta^2)- \alpha(1-{\mathrm i}\beta)\bigr]=G\biggl(\alpha_c-\frac{w}{u}\biggr)\, ,
\end{equation}
where $D\equiv {S_0 C_X}/{\alpha^2}$  and $G\equiv {S_0 C_X(1+\beta^2)}/{\alpha^2}$. We separate the equation into its real and imaginary parts to obtain
\begin{gather}
\omega_{r}^{2}-\omega_{i}^{2}=D\bigl[\alpha_c\left(1+\beta^{2}\right)-\alpha\bigr]\,,\\
2\omega_r\omega_{i}=D\alpha\beta \hskip 8mm \Longrightarrow \hskip 8mm \omega_r=\frac{D\alpha\beta}{2\omega_{i}}\, .
\end{gather}
Insert now (B.4) into (B.3) and define $F\equiv D/2$
and $s\equiv\omega_{i}^{2}$ we obtain a quadratic equation for $s$, namely
\be
 s^{2}+2 F\bigl[\alpha_{c}\left(1+\beta^{2}\right)-\alpha\bigr] s-F^{2}\beta^{2}\alpha^{2}=0\,,
\ee
whose solutions are
\be
s_{1\pm}=\underbrace{-F}_{\mathcal{A}1}\underbrace{\bigl[\alpha_{c}\left(1+\beta^{2}\right)-\alpha\bigr]}_{\mathcal{A}2}\underbrace{\left\{1\pm\left[1+\left(\frac{\beta\alpha}{\alpha_{c}\left(1+\beta^{2}\right)-\alpha}\right)^2\right]^{1/2}\right\}}_{\mathcal{A}3}.
\ee
Hence, in terms of $\omega_{i}$
\be
w_{{\mathrm i}\pm}=\pm\sqrt{-F(\alpha_{c}\left(1+\beta^{2}\right)-\alpha)}\sqrt{1+\left[1+\left(\frac{\beta\alpha}{\alpha_{c}\left(1+\beta^{2}\right)-\alpha}\right)^2\right]^{1/2}}\,.
\ee
Since $s$ and $F$ are positive, (B.6) implies that the product $\mathcal{A}2\times \mathcal{A}3<0$, meaning the equation distinguishes only between neutral and non-neutral modes but not between growing and decaying modes. In order to determine between growth and decay one needs to examine the modal relations between the fields $(u,v,w, \hat\varPi )$, where the growing modes are the ones in which all four fields have the same sign (as mentioned at the end of section 3). Consequently, the condition corresponding to the growing modes is
\be
\alpha_{c}-\frac{\alpha}{1+\beta^{2}}< 0  \hskip 8mm \Longrightarrow\hskip8mm
\frac{\alpha}{\sqrt{1+\beta^2}}=\left | \frac{w}{u}\right |>\alpha_c\sqrt{1+\beta^2}\, .
\ee
The condition in equation (B.8) can be also written as
\be{}
\frac{\alpha}{1+\beta^2}=\operatorname{Re}\left\{\frac{w}{u}\right\}>\alpha_c\, .
\ee{}


\end{document}